# Shock-wave-like emission of spin waves induced by interfacial Dzyaloshinskii-Moriya interaction

Hong Xia [1,2], Haoran Chen [1], Changyeon Won [3], Haibin Zhao [2]* and Yizheng Wu [1,4]*

[1] Department of Physics, State Key Laboratory of Surface Physics, Fudan University, 200433, China

[2] Key Laboratory of Micro and Nano Photonic Structures (Ministry of Education) and Shanghai Ultra-precision Optical Manufacturing Engineering Research Center, Department of Optical Science and Engineering, Fudan University, 200433, China

[3] Department of Physics, Kyung Hee University, Seoul 02447, South Korea

[4] Shanghai Research Center for Quantum Sciences, Shanghai 201315, China

## Abstract

We investigated spin wave (SW) propagation and emission in thin film systems with strong interfacial Dzyaloshinskii-Moriya interaction (DMI) utilizing micromagnetic simulation. The effect of DMI on SW propagation is analogous to the flow of magnetic medium leading to the spin Doppler effect, and a spin-polarized current can enhance or suppress it. It is demonstrated that, for a Doppler velocity exceeding a critical value, a shock-wave-like emission of SWs with a cone-shape emerges from a magnetically irregular point as the cone apex. The cone angle is quantitatively determined by the DMI-induced Doppler velocity. Combining the interfacial DMI and the spin-polarized current, a constant SW emission by a static source is demonstrated, which provides a promising route to efficiently generate SWs with tunable frequency.

# I. Introduction

The research field of magnonics targets the information transport and processing based on the spin waves (SWs), and the effective generation and manipulation of SWs are pivotal to the development of magnon spintronics [1,2]. The nonreciprocal propagation of SWs offers an opportunity to control the magnon flowing to a desirable direction [3]. Recently, the Dzyaloshinskii-Moriya interaction (DMI) induced by the interfacial inversion symmetry breaking in magnetic heterostructures attracted great interests, since DMI is responsible for the emergence of topologically protected spin textures, such as Skyrmion[4-8], anti-Skyrmion [9-13], chiral domain walls [14-17], which can be considered as the effective information carriers in the next-generation data storage and spin logic devices [18-20]. Moreover, DMI can also induce the nonreciprocal propagation of SWs [21-23], which provides a prevailing methodology to determine the DMI in magnetic multilayers by measuring the frequency shifts $\Delta f$ between counterpropagating SWs with wave vectors $+k$ and $-k$ [24-31]. Beside the frequency shift, it was theoretically predicted that the interfacial DMI can induce an asymmetric SW group velocity [21], which was experimentally confirmed recently [32]. The interfacial DMI was also utilized to control the SW propagation properties [18-23]. J. Kim *et al.* theoretically discovered the unidirectional SW power flow and the unidirectional caustic beams from a point-source excitation induced by the interface DMI [33]. Recent experiments revealed that the interfacial DMI can generate the magnon drift currents in heterostructures of yttrium iron garnet and platinum [34]. In the presence of inhomogeneous DMIs, the tunable Snell's law at interfaces of regions with different DMI strengths [35] and the tunable spin-wave band gap in a magnonic crystal [36] have been reported. Those new physical properties related to the nonreciprocal SW dispersion induced by DMI pave the way to their application in future magnonics devices.

The DMI-induced SW nonreciprocity is proportional to the DMI constant [21,23,33]. In most previous studies on DMI, the DMI strength is usually less than 2 mJ/m² [32-34], thus the frequency shift $\Delta f$ for wave vectors $+k$ and $-k$ is much smaller than the SW frequency without DMI. Experimentally, the reported DMI values can be $3.3$ mJ/m$^2$ in the Pt/Co/MgO system [37], and be larger than $6$ mJ/m$^2$ in the Co/Fe/W(110) [38] and CoFeB/Pt(110) systems [39]. In the systems with such strong DMI, the asymmetric SW frequency shift induced by DMI may be comparable or even larger than the SW frequency. However, studying the effect of strong DMI on the SW propagation and generation is still lacking. In this paper, utilizing micromagnetic simulation, we investigated the effect of strong DMI on the SW excitation. We first demonstrate that the DMI-induced nonreciprocal SW dispersion can cause the Doppler effect during the SW propagation, which is analogous to the spin Doppler (SD) effect induced by the spin transfer torque (STT) effect through the spin-polarized current [40,41]. In a magnetic film system with strong DMI above a critic value, a local static DC field pulse can induce the emission of SWs, which contain a propagating wavefront with a cone shape, analogous to the acoustic shock waves in the supersonic wind [42]. The cone angle decreases with the DMI strength, which can be quantitatively understood by the DMI-induced Doppler velocity and the SW dispersion in the film system. Our analysis shows that the combination effect of spin-polarized current and DMI can induce the SW emission from a static DC wave source, which can provide a promising route to generate tunable SWs with great potential in magnonic applications based on SWs [1,43].

**II. Simulation methods**

Computational simulation was performed with the modified code package MuMax3 [44] including the interfacial DMI. Different from the studies on a stripe film in the literatures [21,26,27,45], our simulations were mostly performed in a ferromagnetic (FM) disk, which has the advantage of studying SW propagation along all directions. The disk size was $2 \times 2$ μm$^2$ with a thickness of $1$ nm. The unit cell in the simulation was $1 \times 1 \times 1$ nm$^3$. The typical material parameters of permalloy (Py) were chosen as $\mu_0 M_s = 1$ T, the damping constant $\alpha = 0.01$, and exchange stiffness $A_{ex} = 1.3 \times 10^{-11}$ J/m. No magnetic anisotropy was considered in the simulation, thus an external in-plane field $H$ of $0.01\,\text{T}$ was applied in order to align the magnetization along the field direction. An interfacial DMI was considered in the simulation with the DMI energy expressed as $\varepsilon_{DM} = D\left(m_z \frac{\partial m_x}{\partial x} - m_x \frac{\partial m_z}{\partial x} + m_z \frac{\partial m_y}{\partial y} - m_y \frac{\partial m_z}{\partial y}\right)$ [10,21,46]. Here, $m_x$, $m_y$ and $m_z$ are the components of the unit magnetization, and $D$ is the strength of DMI.

In order to calculate the SD effect induced by the spin-polarized current, we also performed the micromagnetic simulation by adding the STT term to the Landau–Lifshitz–Gilbert equation (LLG) [47,48]. In the simulation, the polarization of electron current, $P$, was chosen as $P = 1$ for simplification.

### III. Results and discussion

In Ref. [21], the asymmetric SW dispersion induced by DMI was studied for the wave vector $k$ perpendicular to the magnetization direction. Moon *et al.* included the dipolar field induced by the SWs, and derived the SW dispersion in a film system with DMI, which is expressed in the following [21]:

$$\omega = \gamma\mu_0 \sqrt{\left(Jk^2 + \frac{M_s}{4} + H\right)\cdot\left(Jk^2 + \frac{3M_s}{4} + H\right) - \frac{e^{-4|k|d}M_s^2(1 + 2e^{2|k|d})}{16}} + \frac{2\gamma}{M_s}kD \quad (1)$$

here, $\gamma$ is the gyromagnetic ratio, $\mu_0$ is permeability of vacuum, $J$ is $\frac{2A_{ex}}{\mu_0 M_s}$, $M_s$ is the saturation magnetization, $k$ is the wave vector of SW, and $d$ is the film thickness.

In general, the dispersion of a wave can be expressed by $\omega = \vec{V}_p \cdot \vec{k}$ with $\vec{V}_p$ as the phase velocity of the wave, thus the SW dispersion in Eq. (1) can be rewritten as $\omega = (\vec{V}_0 + \vec{v}_{DMI}) \cdot \vec{k}$. $\vec{V}_0$ is considered as the intrinsic phase velocity of SW in the system without DMI. $\vec{v}_{DMI}$ can be regarded as the SD velocity induced by DMI with the magnitude $v_{DMI} = \frac{2\gamma}{M_s}D$. The dispersion in Eq. (1) is analogous to the wave dispersion in a medium with a constant flowing speed of $\vec{v}_{DMI}$, and then the observed phase velocity by a fixed observer is the sum of the intrinsic phase velocity and the flowing speed of the medium, which can be considered to have the SD effect.

We performed micromagnetic simulation on the film system containing the DMI with $D = 1 \text{ mJ/m}^2$. The magnetization was aligned along the $+x$ direction by the field $H$. A sinusoidal AC field $\widetilde{H}_z$ along $z$ axis was applied at the disk center in a confined circle region with a diameter of $40 \text{ nm}$. $\widetilde{H}_z$ has a magnitude of $0.01 \text{ T}$ and a frequency of $10 \text{ GHz}$. The AC field excites SWs propagating in all directions. Fig. 1(a) shows a snapshot of simulated $m_z$ distribution at $t = 0.5 \text{ ns}$ after applying $\widetilde{H}_z$. It is clear that the wavefronts of SWs show an elliptical shape, which demonstrates the SD effect induced by the flowing of the medium. The simulated SW propagation is different from the Doppler effect induced by a moving source in a static medium, which should possess a circle-shape wave front with the center position changing with time. For $\vec{M} \parallel +\hat{x}$, the calculated SD velocity $\vec{v}_{DMI}$ in Fig. 1(a) is along the $+y$ direction with an amplitude of $440 \text{ m/s}$. In Ref. [33], Kim *et al.* calculated the SW power flow in a DMI system exerted with an AC field pulse, showing that after the field pulse stops, the center of excited SW ripple drifts with a constant drift velocity. Such SW power drift is induced by the interface DMI, and can also be well understood with the physical picture of the SD effect induced by the flowing of the medium in which the SWs propagate.

In ferromagnets, V. Vlaminck *et al.* [40] demonstrated the SD effect induced by a spin-polarized current. In this pioneering study, the standard adiabatic gradient expression of STT can result in a SW frequency shift of $\Delta\omega_{STT} = -\frac{g\mu_B P}{2eM_s}\vec{j}\cdot\vec{k}$, where $\vec{j}$ is the current density, $g$ is the Lande factor, $\mu_B$ is the Bohr magneton, and $e$ is the unit charge. Such a Doppler shift indicates that the spin-polarized current can induce the flowing of the medium along the current direction respective to the SW with a Doppler velocity of $\vec{u}_{STT} = -\frac{g\mu_B P}{2eM_s}\vec{j}$ [40]. We performed the micromagnetic simulation by adding the STT term to the LLG equation [47,48], and proved that both the spin-polarized current and DMI can induce the SD effect. Fig. 1(b) shows the snapshot of simulated $m_z$ distribution while applying a spin-polarized current along $+y$ direction in the film system with zero DMI. The applied current density is $j = 6.08 \times 10^{12}$ A/m$^2$, which can induce a SD velocity of $u_{STT} = 440$ m/s along the $-y$ direction. Thus, when applying a current with the density of $j = 6.08 \times 10^{12}$ A/m$^2$ in the DMI system with $D = 1$ mJ/m$^2$, due to the opposite SD velocity of $\vec{v}_{DMI}$ and $\vec{u}_{STT}$, the effects from DMI and spin-polarized current cancel each other. As shown in Fig. 1(c), the resulting SWs symmetrically propagate along all directions, similar to that in the system without DMI. If the current direction is switched, the SD effect can be enhanced, as demonstrated by the dashed elliptical circle in Fig. 1(d). Note that, to observe SD effect and get the SW dispersion, only in this part the AC field is used. In the rest of the paper all of the wave sources are set as a DC field.

Next, we demonstrate that the static DC wave source may induce shock-wave-like emission of SWs when the Doppler velocity induced by a strong DMI is large enough. The acoustic shock wave with a cone shape can be found for an aircraft traveling faster than the speed of sound in air, or the supersonic wind in a wind tunnel passing through a static obstacle [42]. In ferromagnets with zero DMI, the shock-wave-like emission of SWs with a Mach cone shape has been theoretically predicted by a fast moving local field, which was named as the spin Cherenkov effect [49]. So, before studying the SW emission induced by DMI, we first simulated the shock-wave-like emission of SWs in the system with zero DMI by moving the source. The diameter of wave source is 10 nm and the strength of the DC field is $H_z = 0.1$ T. The source was moved along the $-y$ direction with a fixed velocity $v_h$. Figs. 2(a-c) show the snapshot images of simulated $m_z$ distribution at $t = 0.5$ ns after moving the magnetic source with velocity $v_h$ of 700, 1400 and 2000 m/s, respectively. For $v_h$<1200 m/s, no SW was observed. If $v_h$ was higher than 1200 m/s, there was a clear SW Mach cone induced by the moving source. For the higher $v_h$, the Mach cone becomes sharper. Fig. 2(d) shows the profiles of the SWs measured along the red dashed line in Fig. 2(b) and the blue dashed line in Fig. 2(c). It is clear that a SW with shorter wavelength and larger amplitude is excited along the $-y$ direction, and there is a weaker SW with longer wavelength along the $+y$ direction.

As demonstrated in Fig. 1, the SD velocity of SWs in the DMI system is equivalent to the flowing velocity of the magnetic medium. If the medium flows across the obstacle with a high enough SD velocity induced by strong DMI, the shock-wave-like emission of SWs is expected. In a DMI system, we applied a DC field of $H_z = 0.1$ T fixed in the film center with a diameter of 10 nm, which can be considered as a magnetic obstacle in the flowing magnetic medium. Figs. 2(e-g) show the snapshot images of simulated $m_z$ distribution in the systems with different DMI strengths of 1.59, 3.18 and 4.55 mJ/m$^2$, which give equivalent Doppler velocities of 700, 1400, and 2000 m/s, respectively. Figs. 2(f) and 2(g) show the SW emission with clear cone shape, similar to the SW emission by the moving source in Figs. 2(b) and 2(c). Fig. 2(h) shows the line profiles of SWs across the dashed lines in Figs. 2(f) and 2(g).

The shock-wave-like emission of SWs with a Mach cone in ferromagnets by a fast moving source has been first investigated by M. Yan *et al*. [49]. Analogous to the acoustic shock wave, the relative motion between the source and the medium can create disturbance of the medium at the front surface of the source. If the relative velocity is higher than the phase velocity of SWs in the medium, the disturbance to the medium cannot propagate out and accumulates at the front surface of the source. The accumulated energy eventually radiates out through the shock-wave-like emission of SWs. Due to the nonlinear dispersion $\omega(k)$ of SWs, the phase velocity $v_p(k)$ has a minimum $v_0$ at a specific nonzero wave vector $k_0$. Only for $k > k_0$, the SW group velocity $v_g(k)$ (=d$\omega$/d$k$) is larger than $v_p(k)$ (= $\omega/k$), and the minimum $v_0$ is determined by $\omega/k =$ d$\omega$/d$k$. The calculated value of $v_0$ is 1204 m/s in our system. The velocity of the moving source has to overcome $v_0$ in order to excite the SWs, and the excited SWs along the relative motion should have a phase velocity $v_p$ equal to $v_h$ due to the velocity match under the resonance condition.

In the thin film with zero DMI, we excited the SWs by an AC field with different frequencies, and determined the $\omega(k)$ dispersion as shown in the inset of Fig. 3(a). Then, the $v_p(k)$ and $v_g(k)$ curves shown in Fig. 3(a) are calculated from the simulated $\omega(k)$ dispersion. The $v_p(k)$ curve with zero DMI is also calculated through Eq. (1). Note that there is a two-fold degeneracy of SWs with a particular $v_p$ above $v_0$, thus, two SWs can be observed for $v_h > v_0$. Both SW modes should have the same $v_p$ equal to $v_h$ [49], and their wave vectors can be determined with the fast Fourier transform of the line profiles along the $y$-axis, thus we can determine the $v_p(k)$ curve through the SW emissions by the local static field. Fig. 3(a) shows that the $v_p(k)$ curve though the SW emission by a moving source agrees well with the calculation from the $\omega(k)$ dispersion.

As indicated by the simulations in Figs. 2(e-g), when the relative speed $v_r$ between the fixed obstacle and the flowing medium induced by DMI is greater than $v_0$, the shock-wave-like emission of SWs occurs. The DMI-induced SD velocity $v_{DMI}$ in our system is larger than $v_0$ for $D > 2.737$ mJ/m$^2$. The excited SWs contain two modes with $v_p$ equal to $v_{DMI}$. By analyzing the line profiles along the symmetric axis, we also obtained the $v_p(k)$ curve, which is the same as that obtained by moving the magnetic sources. Our studies demonstrate that SW emission can be induced by the fast-motion of a local field or the fast flowing of the medium, which is similar to the acoustic shock waves induced by supersonic aircraft or supersonic wind [42].

Fig. 2 shows that the cone angle $\theta_c$ is sharper for larger $v_h$ or $v_{DMI}$. We can quantify $\theta_c$ by the wavefront, as defined in Fig. 2(g). $\theta_c$ is found to be determined by $v_r$, which is equal to $v_h$ for the moving source and $v_{DMI}$ in the DMI system. Fig. 3(b) shows the quantified $\theta_c$ as a function of $v_r$, which well fits with the function of $\theta_c = \sin^{-1}\frac{v_0}{v_r}$. As indicated by the excited SWs in Fig. 3(c), the moving source can excite all the SWs with $v_p(k) \geq v_0$, thus, during the source travelling process, all the SWs with the same $v_p$ excited at different times can form a straight wave front with a cone angle of $\sin^{-1}\frac{v_p}{v_h}$. Note that $v_0$ is the lowest phase velocity of the excited SWs, and we also find that the moving source with the velocity near $v_0$ can excite the SWs with the larger amplitude. Thus, it is reasonable to observe a smallest cone angle of $\theta_c = \sin^{-1}\frac{v_0}{v_h}$. This cone angle of the SW emission is analogous to that in Cherenkov light radiation [50,51], which is equal to $\sin^{-1}\frac{c}{nv}$, with $n$ as the index of refraction, $c$ as the speed of light in vacuum, and *v* as the speed of the moving particle. However, different from the low dispersion of electromagnetic waves, SW dispersion in a magnetic medium shows much stronger nonlinearity (Fig 3(a)), and the SW dispersion is also anisotropic, i.e., it depends on the relative direction between wave vector and magnetization. Thus, it is difficult to calculate all the wavefronts of the emitted SWs.

Although Fig. 2 shows that both fast moving source and the fixed source with strong DMI can induce the SW emission with a Mach cone shape, it should be noted that the formation of the cone shape of SW is not sufficient to evidence the generation of magnetic shock wave. In comparison with the definition of acoustic shock wave [42], one can expect that the magnetic shock wave should have the surface of discontinuity, which can't be identified from Fig. 2. The possible reason may be that the SWs are the transverse waves with the magnetization variation perpendicular to the wave propagation direction, but the acoustic wave in air is the longitudinal wave with the vibrating surface along the propagating direction. The SWs contain much larger dispersion than the acoustic waves, thus the SWs with different wave vectors cannot accumulate their energy at the fixed position to form the discontinuity. However, our analysis shows that the SW emissions in Fig. 2 occur when the group velocity exceeds the phase velocity, analogous to the emission condition of acoustic shock wave. Thus, it is reasonable to name the SW emission with the cone shape in Fig. 2 with the shock-wave-like SW emission.

Next, we discuss the energy origin of the excited SWs. For the SW emission induced by moving sources in Figs. 2(b-c), the SW distributions become unchanged and always follow the motion of the sources. But for the DMI-induced SW emission in Figs. 2(f-g), the system gradually develops into the multi-domain state with chiral domain walls at a few ns after applying the local field pulse. Our simulation shows that the multi-domain state has lower energy than the single-domain state for $D >$ 2.737 mJ/m$^2$ [52], thus the single-domain state is metastable, and the energy of SW excitation is provided by the energy difference while the magnetic system transfers from the single-domain state to the multi-domain state, thus the shock-wave-like SW emission only exists in a limited time range.

The spin-polarized current can induce shock-wave-like SW emission as well, since the Doppler effect induced by the current is analogous to the flowing of the medium [40,41,53]. Fig. 4(a) shows the SW emission induced by a current with $j = 1.93 \times 10^{13}$ A/m$^2$, which can induce a SD velocity of $u_{STT} = 1400$ m/s, but such large current density is unrealistic in a real experimental system. If combining the SD effect induced by DMI and spin-polarized current, the same SW emission can be achieved at lower current density. Fig. 4(b) shows the SW emission in a system with $D = 2$ mJ/m$^2$ and $j = 7.19 \times 10^{12}$ A/m$^2$, and the total SD velocity of magnetic medium is also 1400 m/s, thus the same SW emission can be achieved. Note that in the system with $D = 2$ mJ/m$^2$, the single-domain state is the ground state, thus the energy of SW emission in Fig. 4(b) is from the spin-polarized current, and then a constant SW emission is expected.

In order to demonstrate that constant SW emission can be achieved by combining DMI and spin-polarized current, we performed micromagnetic simulation on a long magnetic strip with a simulation time up to 20 ns. The strip had a length of 8 μm and a width of 200 nm with the periodic boundary widthwise. After the magnetization was aligned along the $+x$ axis by a field of 0.01 T, we applied a local field $H_z$ of 0.1 T in the center region with a width of 10 nm, then performed the simulation with different DMI strength $D$ and current density $j$. Fig. 4(c) shows the magnetization distribution at $t = 10$ ns after applying the local field at the stripe center, and we find the SW configuration is nearly unchanged for $t > 8$ ns, indicating a constant SW source. For $j$ along $+y$ direction, the total SD velocity is only $359.5 \, m/s$, thus no SW emission can be observed. By measuring the wave vectors $k$ of excited SWs, we can determine $f$ of the excited SWs as a function of $j$ and $D$, as shown in Fig. 4(d). There is a minimum current density $j_c$ to induce the SW emission for a fixed DMI strength, and $j_c$ linearly decreases with $D$, as shown in Fig. 4(e). In order to reach the condition for the SW emission, the effective Doppler velocity should satisfy the condition $v_{DMI} + u_{STT} \geq v_0$, thus the minimum current density $j_c$ is equal to $\frac{2e}{g\mu_B P}(M_s v_0 - 2\gamma D)$, which agrees well with the simulation results in Fig. 4(e). For $D = 2$ mJ/m$^2$, the required $j_c$ to induce the SW emission is only $4.48 \times 10^{12}$ A/m$^2$, and it becomes smaller for the stronger $D$.

The predicted shock-wave-like SW emission is possible to be realized in real experimental systems. The reported DMI value can be as large as 3.3 mJ/m$^2$ in the Pt/Co/MgO system [37] and 6.55 mJ/m$^2$ in the Co/Fe bilayer grown on W(110) substrate [38], which is beyond the critic value of DMI to emit the SWs. In principle, SW excitation under spin-polarized current can be measured by BLS [54-57], thus the predicted SW excitation induced by the combined effect of DMI and spin-polarized current is plausible to be verified experimentally. It thereby provides a promising route towards tunable SW generation and valuable DC to AC signal transformation with low current threshold, which has important potential applications in magnonic devices, such as spin-torque oscillators [58,59]. Spin-torque oscillators utilize spin-polarized current to generate the AC signal, and our studies show that the interfacial DMI could greatly reduce the critical current density for SW excitation. Fig. 4(d) shows that SWs excited by the shock-wave-like emission can reach very high frequency up to 100 GHz, which could provide great potential for wide band spintronics applications with the tunable frequency.

## IV. Summary

In summary, utilizing micromagnetic simulation, we demonstrated that the spin Doppler effect induced by interfacial DMI in a thin FM film is analogous to the flow of magnetic medium, and strong DMI or spin-polarized current can induce SW emission at a local magnetic defect for the Doppler velocity higher than the minimum phase velocity of SWs in the system. The SW excitation with the cone shape induced by strong DMI is analogous to the acoustic shock wave induced by supersonic aircraft or supersonic wind. The cone angle of emitted SW wavefronts can be quantitatively calculated from the non-reciprocal SW dispersion and the DMI strength. Combining the effects of spin-polarized current and DMI, constant SW emission can be achieved in thin film systems with SW frequency tuned by the current density. Due to the fundamental interests and potential applications in magnonic devices based on SWs, the predicted constant shock-wave-like SW emission in thin DMI systems deserves further experimental exploration.

**Acknowledgments**

This work in Fudan University was supported by the National Natural Science Foundation of China (Grant No. 11734006, No. 11974079, and No. 11774064), the National Key Research and Development Program of China (Grant No. 2016YFA0300703) and the Shanghai Municipal Science and Technology Major Project (Grant No. 2019SHZDZX01). The work in Kyung Hee University was supported by the NRF, funded by the Korean Government (NRF-2018R1D1A1B07047114).

References:

[1] A. V. Chumak, V. I. Vasyuchka, A. A. Serga and B. Hillebrands, Nat. Phys. **11**, 453 (2015).

[2] H. M. Yu, J. Xiao and H. Schultheiss, Phys. Rep. **905**, 1 (2021).

[3] J. H. Kwon, J. Yoon, P. Deorani, J. M. Lee, J. Sinha, K. J. Lee, M. Hayashi and H. Yang, Sci. Adv. **2**, e1501892 (2016).

[4] A. Fert, N. Reyren and V. Cros, Nat. Rev. Mater. **2**, 17031 (2017).

[5] W. J. Jiang, G. Chen, K. Liu, J. D. Zang, S. G. E. te Velthuis and A. Hoffmann, Phys. Rep. **704**, 1 (2017).

[6] X. Z. Yu, W. Koshibae, Y. Tokunaga, K. Shibata, Y. Taguchi, N. Nagaosa and Y. Tokura, Nature **564**, 95 (2018).

[7] N. Kanazawa, S. Seki and Y. Tokura, Adv. Mater. **29**, 1603227 (2017).

[8] B. S. Kim, J. Phys.: Condens. Matter **31**, 383001 (2019).

[9] W. Koshibae and N. Nagaosa, Nat. Commun. **7**, 10542 (2016).

[10] S. Y. Huang, C. Zhou, G. Chen, H. Y. Shen, A. K. Schmid, K. Liu and Y. Z. Wu, Phys. Rev. B **96**, 144412 (2017).

[11] U. Ritzmann, S. von Malottki, J. V. Kim, S. Heinze, J. Sinova and B. Dupe, Nat. Electron. **1**, 451 (2018).

[12] V. Kumar, N. Kumar, M. Reehuis, J. Gayles, A. S. Sukhanov, A. Hoser, F. Damay, C. Shekhar, P. Adler and C. Felser, Phys. Rev. B **101**, 014424 (2020).

[13] J. Jena, R. Stinshoff, R. Saha, A. K. Srivastava, T. P. Ma, H. Deniz, P. Werner, C. Felser and S. S. P. Parkin, Nano Lett. **20**, 59 (2020).

[14] A. Brataas, Nat. Nanotechnol. **8**, 485 (2013).

[15] G. Chen, S. P. Kang, C. Ophus, A. T. N'Diaye, H. Y. Kwon, R. T. Qiu, C. Won, K. Liu, Y. Z. Wu and A. K. Schmid, Nat. Commun. **8**, 15302 (2017).

[16] R. P. del Real, V. Raposo, E. Martinez and M. Hayashi, Nano Lett. **17**, 1814 (2017).

[17] J. Lucassen, M. J. Meijer, O. Kurnosikov, H. J. M. Swagten, B. Koopmans, R. Lavrijsen, F. Kloodt-Twesten, R. Fromter and R. A. Duine, Phys. Rev. Lett. **123**, 157201 (2019).

[18] A. Hoffmann and S. D. Bader, Phys. Rev. Applied **4**, 047001 (2015).

[19] A. Fert and F. N. Van Dau, C. R. Phys. **20**, 817 (2019).

[20] A. Hirohata, K. Yamada, Y. Nakatani, I. L. Prejbeanu, B. Dieny, P. Pirro and B. Hillebrands, J. Magn. Magn. Mater. **509**, 166711 (2020).

[21] J. H. Moon, S. M. Seo, K. J. Lee, K. W. Kim, J. Ryu, H. W. Lee, R. D. McMichael and M. D. Stiles, Phys. Rev. B **88**, 184404 (2013).

[22] D. Cortes-Ortuno and P. Landeros, J. Phys.: Condens. Matter **25**, 156001 (2013).

[23] L. Udvardi and L. Szunyogh, Phys. Rev. Lett. **102**, 207204 (2009).

[24] M. Belmeguenai, J. P. Adam, Y. Roussigne, S. Eimer, T. Devolder, J. V. Kim, S. M. Cherif, A. Stashkevich and A. Thiaville, Phys. Rev. B **91**, 180405 (2015).

[25] A. K. Chaurasiya, S. Choudhury, J. Sinha and A. Barman, Phys. Rev. Applied **9**, 014008 (2018).

[26] H. T. Nembach, J. M. Shaw, M. Weiler, E. Jue and T. J. Silva, Nat. Phys. **11**, 825 (2015).

[27] J. Cho, N. H. Kim, S. Lee, J. S. Kim, R. Lavrijsen, A. Solignac, Y. X. Yin, D. S. Han, N. J. J. van Hoof, H. J. M. Swagten, B. Koopmans and C. Y. You, Nat. Commun. **6**, 7635 (2015).

[28] K. Di, V. L. Zhang, H. S. Lim, S. C. Ng, M. H. Kuok, J. W. Yu, J. B. Yoon, X. P. Qiu and H. S. Yang, Phys. Rev. Lett. **114**, 047201 (2015).

[29] X. Ma, G. Q. Yu, S. A. Razavi, S. S. Sasaki, X. Li, K. Hao, S. H. Tolbert, K. L. Wang and X. Q. Li,

Phys. Rev. Lett. **119**, 027202 (2017).

[30] J. Lucassen, C. F. Schippers, M. A. Verheijen, P. Fritsch, E. J. Geluk, B. Barcones, R. A. Duine, S. Wurmehl, H. J. M. Swagten, B. Koopmans and R. Lavrijsen, Phys. Rev. B **101**, 064432 (2020).

[31] J. M. Lee, C. Jang, B.-C. Min, S.-W. Lee, K.-J. Lee and J. Chang, Nano Lett. **16**, 62 (2016).

[32] H. C. Wang, J. L. Chen, T. Liu, J. Y. Zhang, K. Baumgaertl, C. Y. Guo, Y. H. Li, C. P. Liu, P. Che, S. Tu, S. Liu, P. Gao, X. F. Han, D. P. Yu, M. Z. Wu, D. Grundler and H. M. Yu, Phys. Rev. Lett. **124**, 027203 (2020).

[33] J. V. Kim, R. L. Stamps and R. E. Camley, Phys. Rev. Lett. **117**, 197204 (2016).

[34] R. Schlitz, S. Velez, A. Kamra, C.-H. Lambert, M. Lammel, S. T. B. Goennenwein and P. Gambardella, Phys. Rev. Lett. **126**, 257201 (2021).

[35] J. Mulkers, B. Van Waeyenberge and M. V. Milosevic, Phys. Rev. B **97**, 104422 (2018).

[36] S. J. Lee, J. H. Moon, H. W. Lee and K. J. Lee, Phys. Rev. B **96**, 184433 (2017).

[37] A. N. Cao, R. Z. Chen, X. R. Wang, X. Y. Zhang, S. Y. Lu, S. S. Yan, B. Koopmans and W. S. Zhao, Nanotechnology **31**, 155705 (2020).

[38] S. Tsurkan and K. Zakeri, Phys. Rev. B **102**, 060406 (2020).

[39] C. Q. Liu, Y. B. Zhang, G. Z. Chai and Y. Z. Wu, Appl. Phys. Lett. **118**, 262410 (2021).

[40] V. Vlaminck and M. Bailleul, Science **322**, 410 (2008).

[41] R. D. McMichael and M. D. Stiles, Science **322**, 386 (2008).

[42] L. D. Landau and E. M. Lifshitz, in *Fluid Mechanics-2nd ed* (Pergamon Press, Oxdord, 1987), pp. 82-96.

[43] V. V. Kruglyak, S. O. Demokritov and D. Grundler, J. Phys. D: Appl. Phys. **43**, 264001 (2010).

[44] A. Vansteenkiste, J. Leliaert, M. Dvornik, M. Helsen, F. Garcia-Sanchez and B. Van Waeyenberge, AIP Adv. **4**, 107133 (2014).

[45] S. Seki, Y. Okamura, K. Kondou, K. Shibata, M. Kubota, R. Takagi, F. Kagawa, M. Kawasaki, G. Tatara, Y. Otani and Y. Tokura, Phys. Rev. B **93**, 235131 (2016).

[46] J. Xia, X. C. Zhang, M. Yan, W. S. Zhao and Y. Zhou, Sci. Rep. **6**, 25189 (2016).

[47] S. Zhang and Z. Li, Phys. Rev. Lett. **93**, 127204 (2004).

[48] A. Thiaville, Y. Nakatani, J. Miltat and Y. Suzuki, Europhys. Lett. **69**, 990 (2005).

[49] M. Yan, A. Kakay, C. Andreas and R. Hertel, Phys. Rev. B **88**, 220412 (2013).

[50] P. A. Cerenkov, Phys. Rev. **52**, 0378 (1937).

[51] J. V. Jelley, Brit. J. Appl. Phys. **6**, 227 (1955).

[52] See Supplemental Materials, for details on the time-dependent SW emissions, the total energy of the systems with different DMI strength and the temperature effect on the SW generations.

[53] N. I. Polushkin, Appl. Phys. Lett. **99**, 182502 (2011).

[54] M. Madami, S. Bonetti, G. Consolo, S. Tacchi, G. Carlotti, G. Gubbiotti, F. B. Mancoff, M. A. Yar and J. Akerman, Nat. Nanotechnol. **6**, 635 (2011).

[55] K. An, D. R. Birt, C.-F. Pai, K. Olsson, D. C. Ralph, R. A. Buhrman and X. Li, Phys. Rev. B **89**, 140405 (2014).

[56] M. Madami, E. Iacocca, S. Sani, G. Gubbiotti, S. Tacchi, R. K. Dumas, J. Akerman and G. Carlotti, Phys. Rev. B **92**, 024403 (2015).

[57] V. Lauer, D. A. Bozhko, T. Bracher, P. Pirro, V. I. Vasyuchka, A. A. Serga, M. B. Jungfleisch, M. Agrawal, Y. V. Kobljanskyj, G. A. Melkov, C. Dubs, B. Hillebrands and A. V. Chumak, Appl. Phys. Lett. **108**, 012402 (2016).

[58] H. Fulara, M. Zahedinejad, R. Khymyn, A. A. Awad, S. Muralidhar, M. Dvornik and J. Akerman,

Sci. Adv. **5**, eaax8467 (2019).

[59] A. Houshang, E. Iacocca, P. Durrenfeld, S. R. Sani, J. Akerman and R. K. Dumas, Nat. Nanotechnol. **11**, 280 (2016).
15

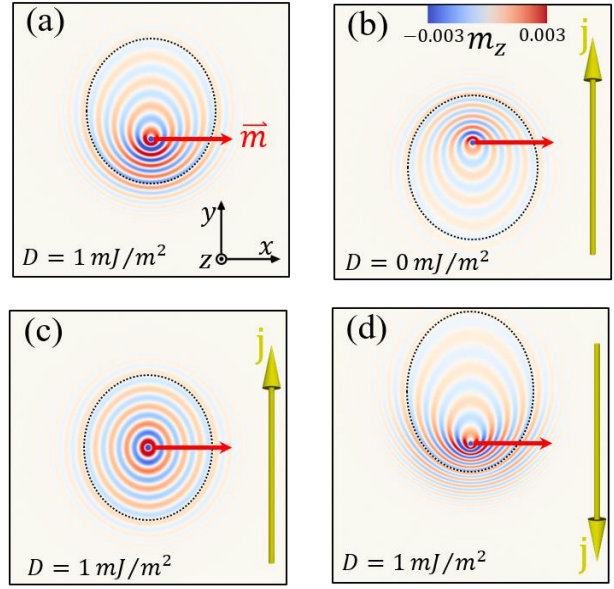

Fig. 1. Snapshot images of simulated $m_z$ distributions at $t = 0.5$ ns after applying a sinusoidal AC field at the image center in the systems with different DMI strength and current density: (a) $D = 1$ mJ/m², $j = 0$ A/m²; (b) $D = 0$ mJ/m², $j = +6.08 \times 10^{12}$ A/m²; (c) $D = 1$ mJ/m², $j = +6.08 \times 10^{12}$ A/m²; (d) $D = 1$ mJ/m², $j = -6.08 \times 10^{12}$ A/m². The frequency of the exciting AC field is 10 GHz. The presented image size is $1 \times 1$ $\mu$m². The red arrows represent the magnetization direction. The black elliptical dash lines show the wavefront and indicate the different Doppler effects.

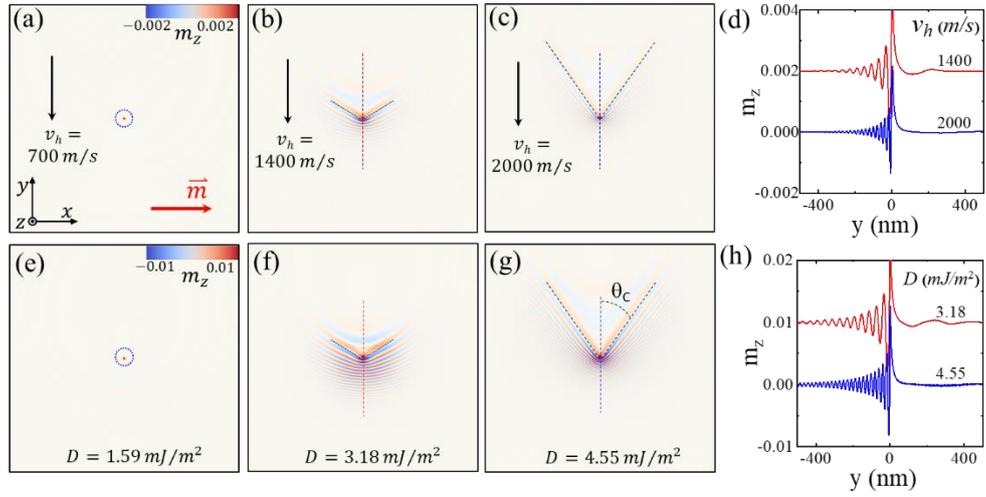

Fig. 2. Shock-wave-like SW emissions. (a-c) Snapshot images of simulated $m_z$ distribution after applying moving sources with a local field $H_z = 0.1$ T in a circle with a diameter of 10 nm. The sources are moving along the dashed arrows with a speed of (a) 700 m/s, (b) 1400 m/s and (c) 2000 m/s. (d) Line profiles of the SWs along the dashed lines in (b) and (c) respectively. (e-g) Snapshot images of $m_z$ distribution with different DMIs: (e) $D = 1.59$ mJ/m², (f) $D = 3.18$ mJ/m², and (g) $D = 4.55$ mJ/m². (h) Line profiles of the SWs along the dashed lines in (f) and (g). The dashed circles in (a) and (d) highlight the location of the local field.

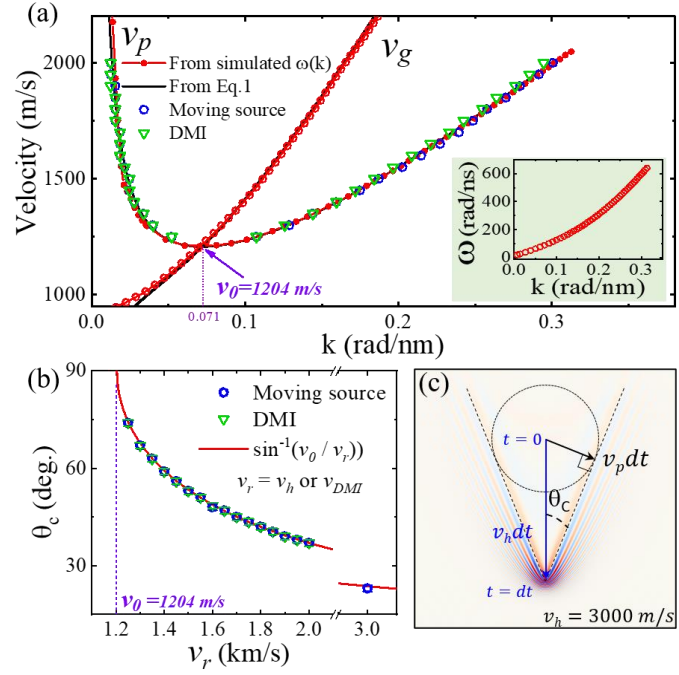

Fig. 3. (a) Calculated phase velocity $v_p(k)$ and group velocity $v_g(k)$ of SWs. The red circles for $v_p(k)$ and $v_g(k)$ are extracted from the SW dispersion $\omega(k)$ with zero DMI shown in the inset. The black solid lines for $v_p(k)$ and $v_g(k)$ are calculated from the theoretical $\omega(k)$ dispersion in Eq. 1. The blue circles and the green triangles are extracted from the simulations with a moving field pulse and DMI, respectively. (b) Mach cone angle $\theta_c$ of SWs as a function of the relative velocity $v_r$ between the source and the medium. (c) Explanation of Mach cone angle $\theta_c$. The SW emission in (c) is induced by the moving local field with a speed of 3000 m/s.

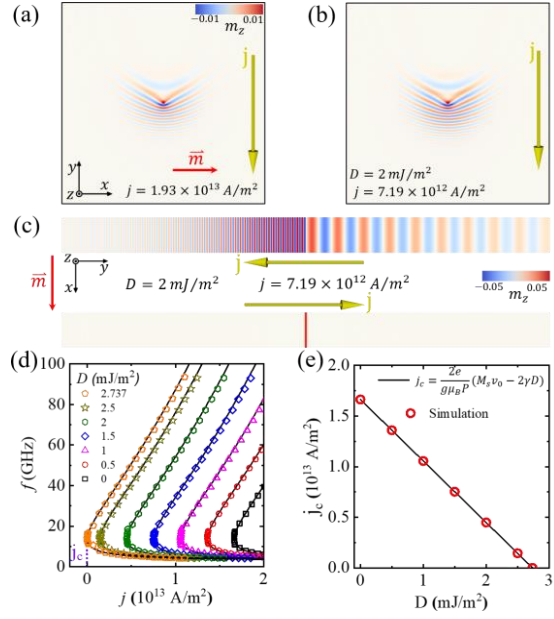

Fig. 4. (a-b) Snapshot images of simulated $m_z$ distribution after applying a local field with (a) $D = 0$ mJ/m², $j = 1.93 \times 10^{13}$ A/m², and (b) $D = 2$ mJ/m², $j = 7.19 \times 10^{12}$ A/m². (c) Snapshot images in a magnetic strip with $D = 2$ mJ/m² and $j = 7.19 \times 10^{12}$ A/m² with opposite current directions. The presented strip has a length of $4\ \mu$m and a width of 200 nm. The snapshot time is $t = 10\ ns$. (d) Excited SW frequency as a function of $j$ with different $D$. (e) Minimum current density $j_c$ as a function of $D$. The black line is the calculated result, and the red circles are from the simulation.

# Supplemental information

# Shock-wave-like emission of spin waves induced by interfacial Dzyaloshinskii-Moriya interaction

Hong Xia [1,2], Haoran Chen [1], Changyeon Won [3], Haibin Zhao [2]* and Yizheng Wu [1,4]*

[1] Department of Physics, State Key Laboratory of Surface Physics, Fudan University, 200433, China

[2] Key Laboratory of Micro and Nano Photonic Structures (Ministry of Education) and Shanghai Ultra-precision Optical Manufacturing Engineering Research Center, Department of Optical Science and Engineering, Fudan University, 200433, China

[3] Department of Physics, Kyung Hee University, Seoul 02447, South Korea

[4] Shanghai Research Center for Quantum Sciences, Shanghai 201315, China

**1. Time-dependent shock-wave-like emission of spin waves**

In the main text, we show the spin wave (SW) patterns excited either by fast-moving source, or by DMI at the fixed time after the excitation of local field pulse. In Fig. S1, we further show the excited SW patterns at different times after applying the local field pulse. The SW patterns excited by two different methods are almost the same at all the time, which further indicates that the SW excitations by these two methods are nearly equivalent. As mentioned in the main text, the Doppler effect induced by DMI is due to the flowing of the medium in which the SWs propagate, so the moving of the source or the flowing of the medium can generate the same Doppler effect of SWs.

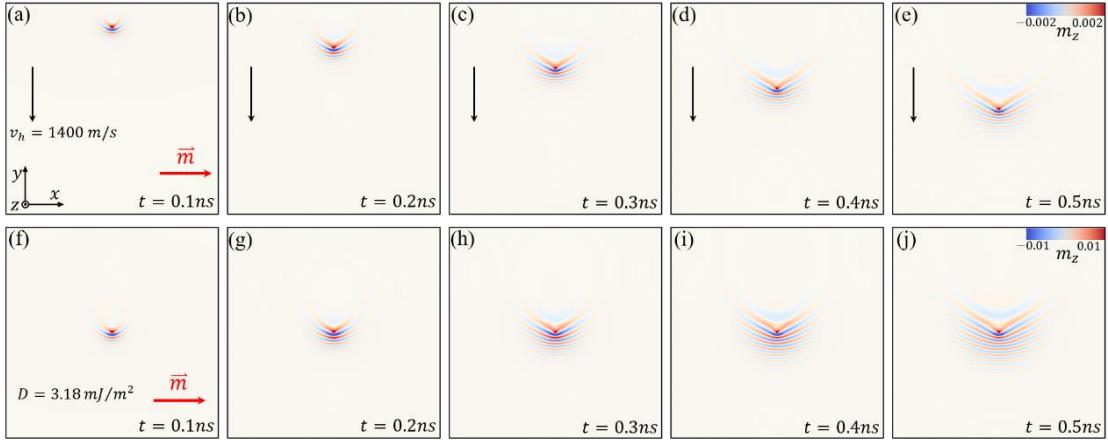

**Figure S1**. The snapshots of the spin wave (SW) emission at different times. (a-e) The SWs are excited by a DC local field moving along the $-y$ axis with the velocity of $1400\ m/s$. The applied local field has the strength of is $H_z = 0.1\ T$ and it is constrained in a circle with the diameter $10\ nm$. (f-j) The SWs are excited by a DC local field at the disk center with the diameter of $10\ nm$ in the present of interfacial Dzyaloshinskii-Moriya interaction (DMI). The applied field is $H_z = 0.1\ T$, and DMI in this calculation is $D = 3.18\ mJ/m^2$. The DMI induced Doppler velocity of the disk is $v_{DMI} = 1400\ m/s$. The formation process of the SW Mach cone in Fig. (f-j) is similar to that in Fig. (a-e). In the simulation, the initial magnetization is along $x$ axis and the size of the disk is $x \times y \times z = 2\ um \times 2\ um \times 1\ nm$.

## 2. The ground state of the system as a function of DMI

In the main text, we discussed the energy origin of the shock-wave-like SW emission induced by strong DMI, and concluded that the energy of SW excitation is provided by the energy difference while the magnetic system transfers from the single-domain state to the chiral multi-domain state. Here, we calculated the ground state energy of the system as a function of the DMI strength. We found that, when the DMI strength $D$ is larger than $2.71\ mJ/m^2$, the ground state of the system changes from a uniform state to a chiral multi-domain state. This critical value is very close to the minimum value of $2.737\ mJ/m^2$ for the DMI induced shock-wave-like emission of SWs. Fig. S2(c) shows the calculated energy of ground states as a function of $D$, which shows that the chiral state becomes the ground state for $D > 2.71\ mJ/m^2$. In order to better show the state transition at $D \sim 2.71\ mJ/m^2$, we further plot the calculated $D$- dependent $R = log_{10}(|\frac{E}{E_u}|)$ in Fig. S2(d), which can better display the transition at $D \sim 2.71 \, mJ/m^2$.

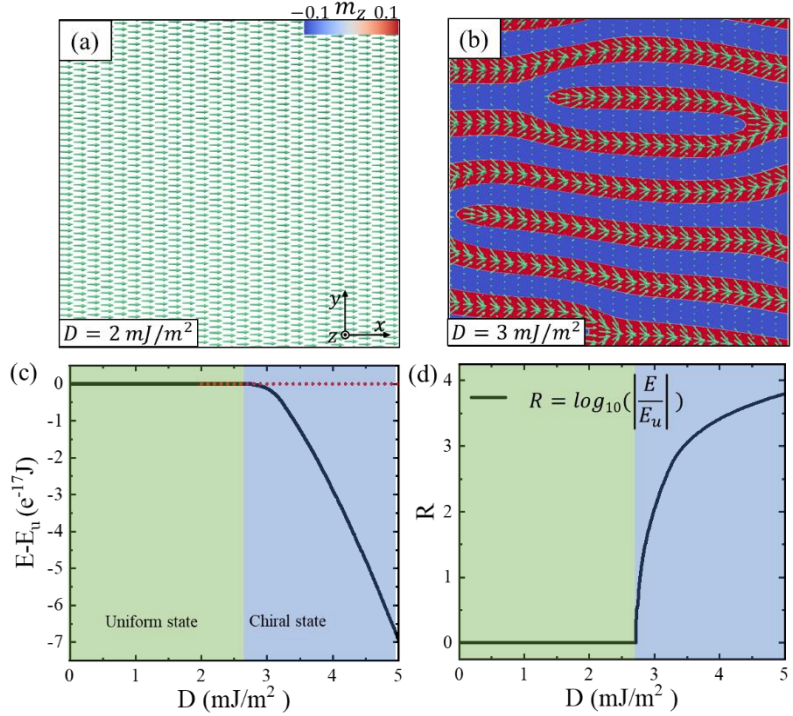

**Figure S2**. The simulated final state relaxed from a random spin state with (a) $D = 2 \, mJ/m^2$ and (b) $D = 3 \, mJ/m^2$. The uniform state is obtained for $D = 2 \, mJ/m^2$, and the multi-domain state with the chiral domain wall is obtained for $D = 3 \, mJ/m^2$. (c) The total energy $E$ of the final state with different DMI strength. For $D < 2.71 \, mJ/m^2$, we find the final state is the uniform state. (d) The calculated DMI-dependent $R = log_{10}(|\frac{E}{E_u}|)$ clearly shows the transition at $D = 2.71 \, mJ/m^2$. $E$ is the energy of the final state, and $E_u$ is the energy of the uniform domain state.

### 3. The thermal effect on the SW emission induced by strong DMI.

All the micromagnetic simulations in the main text were performed at zero temperature. In order to check whether the shock-wave-like SW emission induced by strong DMI is plausible to be observed at room temperature, which is important for further experimental verification of our theoretical studies, we also performed the simulation at finite temperatures.

In our simulation, the temperature effect is simulated by adding the stochastic thermal field $\vec{H}_{therm}$ according to the works by Brown *et al*. [1-5]:

$$\vec{H}_{therm} = \vec{\eta}(step)\sqrt{\frac{2\alpha k_B T}{\gamma \mu_0 M_s \Delta V \Delta T}} \quad (1)$$

where the damping constant $\alpha = 0.01$, the saturation magnetization $\mu_0 M_s = 1$ T, $\gamma$ is the gyromagnetic ratio, $k_B$ is the Boltzmann constant, T is the temperature, the cell volume $\Delta V = 1\ nm^3$, the time step is fixed at $\Delta t = 1 \times 10^{-15} s$, and $\vec{\eta}(step)$ is a random vector from a standard normal distribution whose value is changed after every time step.

Fig. S3(a) shows the simulated SW patterns at different temperatures in the system with the DMI strength $D = 2\ mJ/m^2$ and the spin-polarized current density $j = 7.19 \times 10^{12} A/m^2$. All the parameters are same as those for the calculation of Fig. 4 in the main text. It is clear that the regular SW pattern can only exist for the temperature lower than 100 K. At higher temperature, only random excitation can be observed. However, this phenomenon also indicates the shock-wave-like SW excitation at high temperature. It should be noted that the SWs can be excited not only by the applied local field at the center of stripe, but also by the stochastic thermal field which is randomly distributed. As shown in Fig. 2, the local field can excite the SWs under strong DMI, and the stochastic thermal fields at high temperatures just randomly excite the SWs, which makes the SW patterns irregular and smears out the regular SW pattern induced by the fixed local field in the center.

The SW excitation due to the excitation by DMI and spin-polarized current can further reduce the net magnetization at finite temperatures. It is well known that the net magnetization decreases with the temperature due to the thermal excitation. However, as shown in Fig. S3(b), when we apply DMI and spin-polarized current simultaneously, the net magnetization decreases faster with the temperature than that without DMI, indicating the existence of strong SW excitation. We also calculated the net magnetization as a function of excitation current density $j$ at 300 K in the system with $D = 2\ mJ/m^2$. Fig. S3(c) shows a clear transition of net magnetization at $j = 6 \times 10^{12}\ A/m^2$, which is consistent with the critic current density for the shock-wave-like SW excitation shown in Fig. 4(c) in the main text.

We believe that such SW excitation at room temperature can be detected by the Brillouin light scattering (BLS). As discussed in the main text, the excited SW by the Doppler effect should have the wave vector $k$ satisfying the condition that the phase velocity $v_p(k)$ is equal to the Doppler velocity. The wave vector $k$ of excited SWs should change with the applied current density. So, in a thin film system with strong DMI, it is possible to observe the strong SW intensity at two particular wave vectors $k$, which can be confirmed by the BLS measurement. Moreover, the Doppler velocity can be changed by the applied spin-polarized current, thus the shock-wave-like SW excitation can be further confirmed by the current-dependent BLS measurement. The BLS measurement is, however, beyond our research capability, and our studies call for the further experimental study to confirm our theoretical prediction on the SW excitation due to strong DMIs.

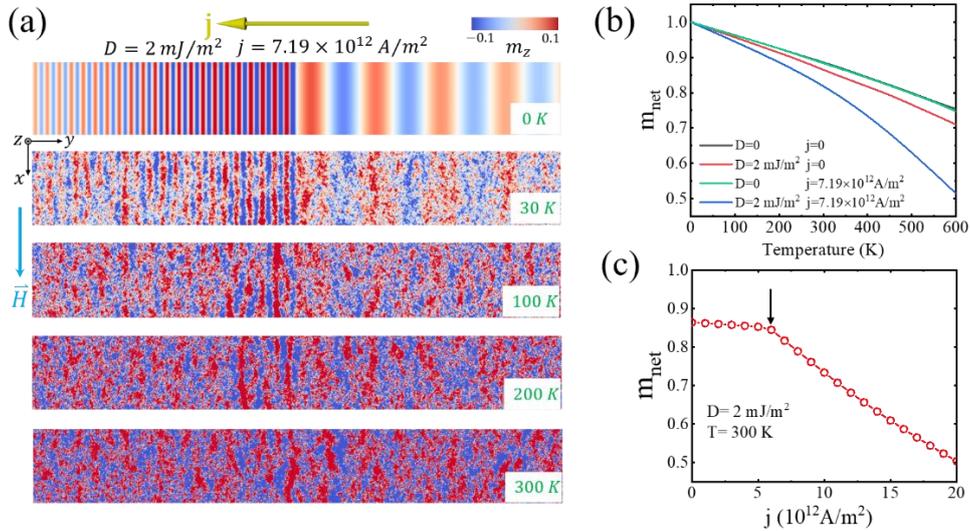

**Figure S3.** The temperature effect on the SW excitation. (a) The simulated SW patterns at different temperatures with $D = 2 mJ/m^2$ and $j = 7.19 \times 10^{12} A/m^2$. (b) The simulated net magnetization as a function of temperature. It is clear that the net magnetization with $D = 2 mJ/m^2$ and $j = 7.19 \times 10^{12} A/m^2$ at finite temperature is much lower due to the SW excitation. (c) The net magnetization as a function of spin-polarized current density at 300 K with $D = 2 mJ/m^2$, and a clear transition can be observed as marked by the arrow.

In the following, we also present the videos of micromagnetic simulation, which can better demonstrate the shock-wave-like emission of SWs induced by strong DMI.

**Video S1**. The shock-wave-like emission of SWs induced by the fast-moving source. A DC local field of $H_z = 0.1\,T$ is applied in a circle with a diameter of $10\,nm$. The velocity of the source is $1400\,m/s$. The animation time is $1\,ns$.

**Video S2.** The shock-wave-like emission of SWs induced by DMI. A local DC field of $H_z = 0.1\,T$ is applied in the center with a diameter of $10\,nm$. The strength of DMI is $D = 3.18\,mJ/m^2$, which induces the Doppler velocity of $v_{DMI} = 1400\,m/s$. The animation time is $1\,ns$.

**Video S3**. Time dependent variation of local magnetizations at the wave source in the present of DMI. The strength of DMI is $D = 2\,mJ/m^2$ with the Doppler velocity of $v_{DMI} = 879.8\,m/s$. The animation time is $1\,ns$ after applying a local DC field of $H_z = 0.1\,T$. At this condition, no SW emission can be observed.

**Video S4.** Time dependent variation of local magnetizations at the wave source in the present of DMI. The strength of DMI is $D = 3.18\,mJ/m^2$ with the Doppler velocity of $v_{DMI} = 1400m/s$. The animation time is $1\,ns$ after applying a local DC field of $H_z = 0.1\,T$. At this condition, the shock-wave-like emission of SWs can be observed, but the magnetization at the wave source shows no oscillation.

**Video S5**. Time dependent variation of local magnetizations at the wave source in the present of strong spin-polarized current of $j = 1.93 \times 10^{13}\,A/m^2$, which induces the Doppler velocity of $u_{STT} = 1400\,m/s$. The animation time is $1\,ns$ after applying a local DC field of $H_z = 0.1\,T$. At this condition, the shock-wave-like emission of SWs can be observed.

**Video S6**. Time dependent variation of local magnetizations at the wave source in the present of strong spin-polarized current of $j = 7.19 \times 10^{12}\,A/m^2$ and DMI strength of $D = 2\,mJ/m^2$. The total Doppler velocity of the medium is $v = v_{DMI} + u_{STT} = 1400 \, m/s$. The animation time is $1 \, ns$ after applying a local DC field of $H_z = 0.1 \, T$. At this condition, the shock-wave-like emission of SWs can be observed.

**Video S7.** Time dependent SW emission in a magnetic stripe with the spin-polarized current of $j = 7.19 \times 10^{13} \, A/m^2$ and DMI of $D = 2 \, mJ/m^2$. The size of the film is $x \times y \times z = 0.2 \, um \times 8 \, um \times 1 \, nm$. A local DC field of $H_z = 0.1 \, T$ is applied in the center with the width of $10 \, nm$. The wave source is a DC magnetic field perpendicular to the film with $H_z = 0.1 \, T$. The total animation time is $100 \, ns$ after applying the local field, which can demonstrate the constant SW emission induced by the combination of spin-polarized current and DMI at this condition.

References:

[1]   W. F. Brown, Phys. Rev. **130**, 1677 (1963).
[2]   D. M. Apalkov and P. B. Visscher, Phys. Rev. B **72**, 4, 180405 (2005).
[3]   L. Breth, D. Suess, C. Vogler, B. Bergmair, M. Fuger, R. Heer and H. Brueckl, J. Appl. Phys. **112**, 4, 023903 (2012).
[4]   A. Vansteenkiste, J. Leliaert, M. Dvornik, M. Helsen, F. Garcia-Sanchez and B. Van Waeyenberge, AIP Adv. **4**, 107133 (2014).
[5]   R. Tomasello, E. Martinez, R. Zivieri, L. Torres, M. Carpentieri and G. Finocchio, Sci. Rep. **4**, 7, 6784 (2014).